\documentclass[a4paper,11pt]{article}
\usepackage{pos}

\usepackage[switch]{lineno}

\title{\vspace{-0.2cm} The Compton-Pair telescope: A prototype for a next-generation MeV $\gamma$-ray observatory}
 \ShortTitle{The ComPair MeV $\gamma$-ray telescope}

\author*[a,b]{Janeth Valverde}

\author[h]{N. Kirschner}
\author[k,d,b]{Z. Metzler}
\author[k]{L. Smith}
\author[a,d]{N. Cannady}
\author[b]{R. Caputo}
\author[b]{C. A. Kierans}
\author[b,j]{I. Liceaga-Indart}
\author[d,k]{A. Moiseev}
\author[l]{L. Parker}
\author[d,k]{M. Sasaki}
\author[b,d]{A. Schoenwald}
\author[f]{D. Shy}
\author[j,b,d]{S. Wasti}
\author[f]{R. Woolf}
\author[c]{A. Bolotnikov}
\author[c]{G. A. Carini}
\author[m]{A. W. Crosier}
\author[m]{T. Caligure}
\author[c]{A. Dellapenna Jr}
\author[c]{J. Fried}
\author[j,b,d]{P. Ghosh}
\author[e]{S. Griffin}
\author[f]{J. E. Grove}
\author[b]{E. Hays}
\author[c]{S. Herrmann}
\author[i]{E. Kong}
\author[b]{J. McEnery}
\author[b]{J. Mitchell}
\author[b]{J. Perkins}
\author[f]{B. Phlips}
\author[f]{C. Sleator}
\author[f]{E. Wulf}
\author[d,a]{A. Zajczyk}

\affiliation[a]{University of Maryland Baltimore County, Baltimore, MD 21250, USA}
\affiliation[b]{NASA Goddard Space Flight Center, Greenbelt, MD 20771, USA}

\affiliation[c]{Brookhaven National Laboratory, USA} 
\affiliation[d]{Center for Research and Exploration in Space Science and Technology, Greenbelt, MD 20771, USA} 
\affiliation[e]{Wisconsin IceCube Particle Astrophysics Center, USA} 
\affiliation[f]{U.S. Naval Research Laboratory, Washington, DC 20375 USA}
\affiliation[h]{George Washington University, DC 20052, USA}  
\affiliation[i]{Technology Service Corporation, Arlington, VA, 22202, USA}
\affiliation[j]{The Catholic University of America, DC 20064, USA} 
\affiliation[k]{University of Maryland, MD 20742, USA} 
\affiliation[l]{Los Alamos National Laboratory, New Mexico 87545, USA} 
\affiliation[m]{Naval Research Enterprise Internship Program, resident at U.S. Naval Research Laboratory, Washington, DC 20375 USA}

\emailAdd{janeth@umbc.edu}

\abstract{The Compton Pair (ComPair) telescope is a prototype that aims to develop the necessary technologies for future medium energy gamma-ray missions and to design, build, and test the prototype in a gamma-ray beam and balloon flight. The ComPair team has built an instrument that consists of 4 detector subsystems: a double-sided silicon strip detector Tracker, a novel high-resolution virtual Frisch-grid cadmium zinc telluride Calorimeter, and a high-energy hodoscopic cesium iodide Calorimeter, all of which are surrounded by a plastic scintillator anti-coincidence detector. These subsystems together detect and characterize photons via Compton scattering and pair production, enable a veto of cosmic rays, and are a proof-of-concept for a space telescope with the same architecture. A future medium-energy gamma-ray mission enabled through ComPair will address many questions posed in the Astro2020 Decadal survey in both the New Messengers and New Physics and the Cosmic Ecosystems themes. In this contribution, we will give an overview of the ComPair project and steps forward to the balloon flight.}

\ConferenceLogo{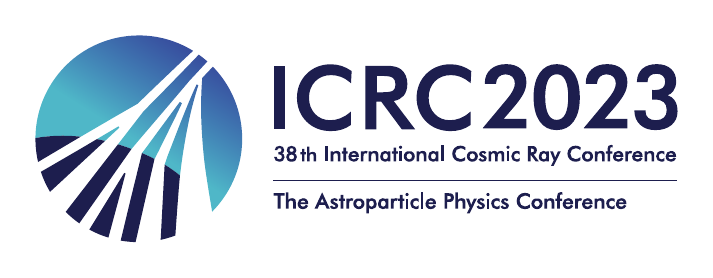}

\FullConference{%
38th International Cosmic Ray Conference (ICRC2023)\\
  26 July - 3 August, 2023\\
  Nagoya, Japan}

\begin{document}
\maketitle

\section{Introduction}

Gamma-ray astronomy is the study of the most energetic form of light and has been yielding important discoveries about our universe since the 1960s, and more recently in multi-messenger astrophysics. For instance, the combined detection of gamma rays with neutrinos from supernova SN1987A \citep{1988Natur.331..416M}, most recently, gravitational waves (GWs) from the binary neutron star merger GW 170817 \citep{2018ApJ...861...85A}, and the spatial correlation with the neutrino-producing blazar TXS 0506+056 \citep{IceCube:2018cha}. 
However, the MeV domain remains largely under explored, with the last major instrument sensitive to that energy range (0.8 -- 30 MeV), the Imaging Compton Telescope \citep[COMPTEL;][]{1993ApJS...86..657S}, de-orbited in 2000. 
The Compton Pair Telescope (ComPair) team is developing the necessary technologies to enable a future MeV mission, like the All-sky Medium Energy Gamma-ray Observatory \citep[AMEGO;][]{kierans20}, which is a mission concept designed to address the 
MeV gap in sensitivity. 

 ComPair is lead out of NASA Goddard Space Flight Center (GSFC), with collaborators at the Naval Research Laboratory (NRL), Los Alamos National Laboratory (LANL), and Brookhaven National Laboratory (BNL). The team has designed, built and tested a prototype of AMEGO in a beam test and is currently preparing for its first balloon flight in August 2023. In the following, we provide an overview of the motivation, mission concept, and current status of ComPair.

\section{Motivation and mission concept}

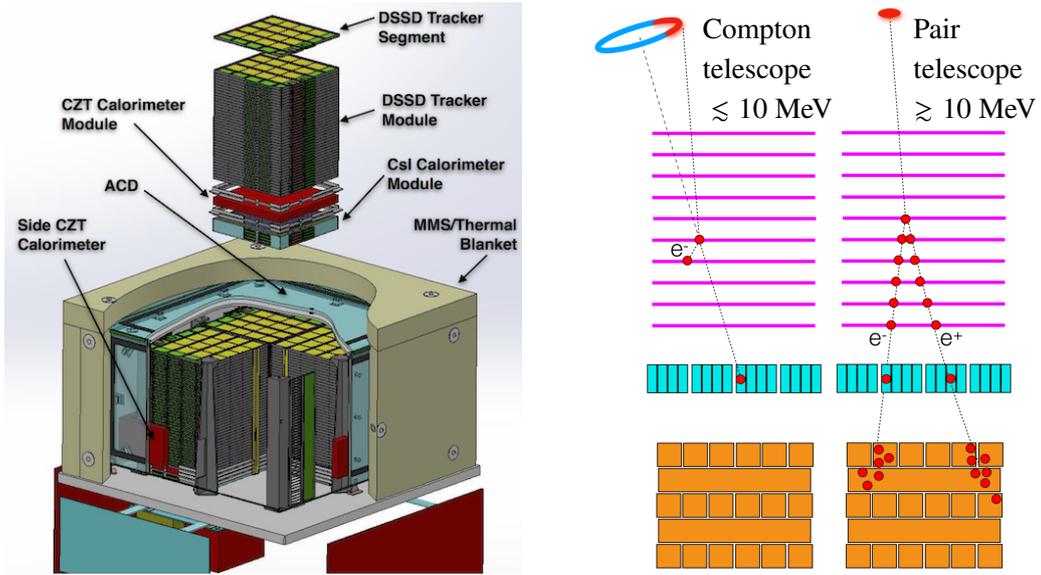
\begin{figure}[h]
\begin{center}
\includegraphics[width=0.45\linewidth,angle=0]{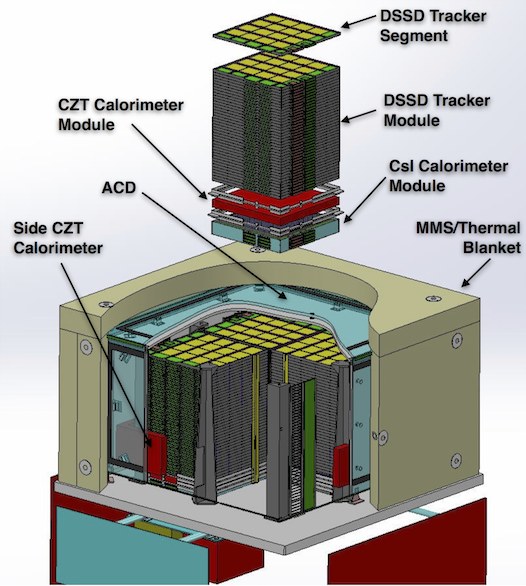}\hspace{8mm}
\includegraphics[width=0.37\linewidth,angle=0]{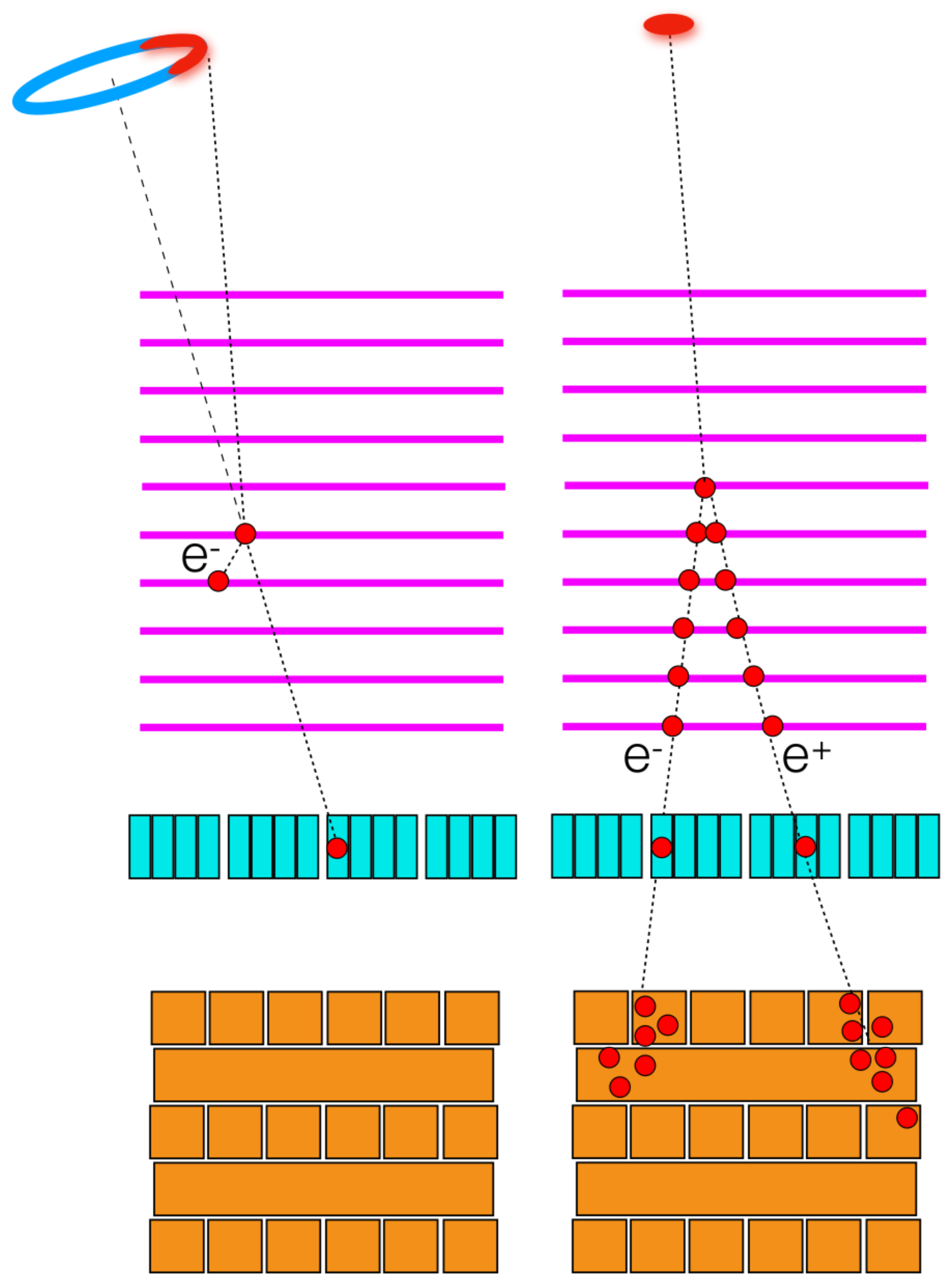}
\put(-117,202){Compton} \put(-117,187){telescope} \put(-117,172){ $\lesssim 10$ MeV}
\put(-38,202){Pair} \put(-38,187){telescope} \put(-38,172){$\gtrsim 10$ MeV}
\caption{\textit{Left:} Schematic view of the AMEGO mission concept, which consists of 60 tracker layers sitting on top of a low energy CZT calorimeter, and the high energy CsI calorimeter is located at the bottom. All of these are surrounded by a plastic ACD. \textit{Right:} With the same detector systems as AMEGO, ComPair is sensitive to the 0.2--25 MeV band, which covers both the Compton and pair conversion regimes. The figure shows 10 of the ComPair tracker layers in magenta, the CZT calorimeter in teal, and the 5 layers of the CsI calorimeter in orange. Compton scattering dominates below $\sim 10$ MeV (shown on the left), where a photon's direction is constrained by the initial Compton scattering angle calculated by the interaction positions and energies. Above $\sim 10$ MeV, pair production dominates (shown on the right) and the initial photon direction is found by tracing the path of the electron and positron pair products in the tracker.  
\label{fig:amego}}
\end{center}
\end{figure}

AMEGO\footnote{\url{https://asd.gsfc.nasa.gov/amego/science.html}} targets the study of astrophysical objects that produce GWs and neutrinos in order to understand the processes occurring under the extreme conditions around compact objects. AMEGO has the goal of deciphering the underlying processes in the jets of gamma-ray bursts and active galactic nuclei, such as blazars. It aims to resolve the processes of element formation in objects such as kilonovae or supernovae \citep{mcenery19}. 
All of these
require an MeV gamma-ray mission with capabilities exceeding those of existing and planned observatories such as Fermi \citep{2009ApJ...697.1071A}, Swift \citep{2005SSRv..120..165B} and COSI \citep{2019BAAS...51g..98T}. 

AMEGO was an MeV mission concept submitted to the Astro2020 Decadal survey, designed to be sensitive to photons from 0.2 MeV to 10 GeV. AMEGO covers both the Compton and pair-conversion regime. To accomplish the detection and imaging of both these processes, AMEGO aims to employ a unique combination of a Double Sided Silicon Strip Detector (DSSD) tracker with a novel high-resolution virtual Frisch-grid Cadmium Zinc Telluride (CZT) low-energy calorimeter, and a thick Cesium Iodide (CsI) high-energy hodoscopic calorimeter that are all surrounded by an anti-coincidence detector (ACD) that rejects cosmic-ray background events, as shown in Fig. \ref{fig:amego}. 

The tracker converts or scatters incoming gamma rays and accurately measures the positions and energies of either the electron-positron pairs or the Compton-scattered electrons passing through the instrument. The low energy calorimeter provides a precise measurement of the location and energy of the low energy scattered gamma rays. The high energy calorimeter is based on the  {\sl Fermi}-LAT CsI calorimeter design and enhances upper energy range measurements by measuring the pair-conversion products. The anti-coincidence detector vetoes cosmic-ray background.

\section{The Compton Pair Telescope}

\begin{figure}[h]
\begin{center}
\includegraphics[width=0.55\linewidth,angle=0]{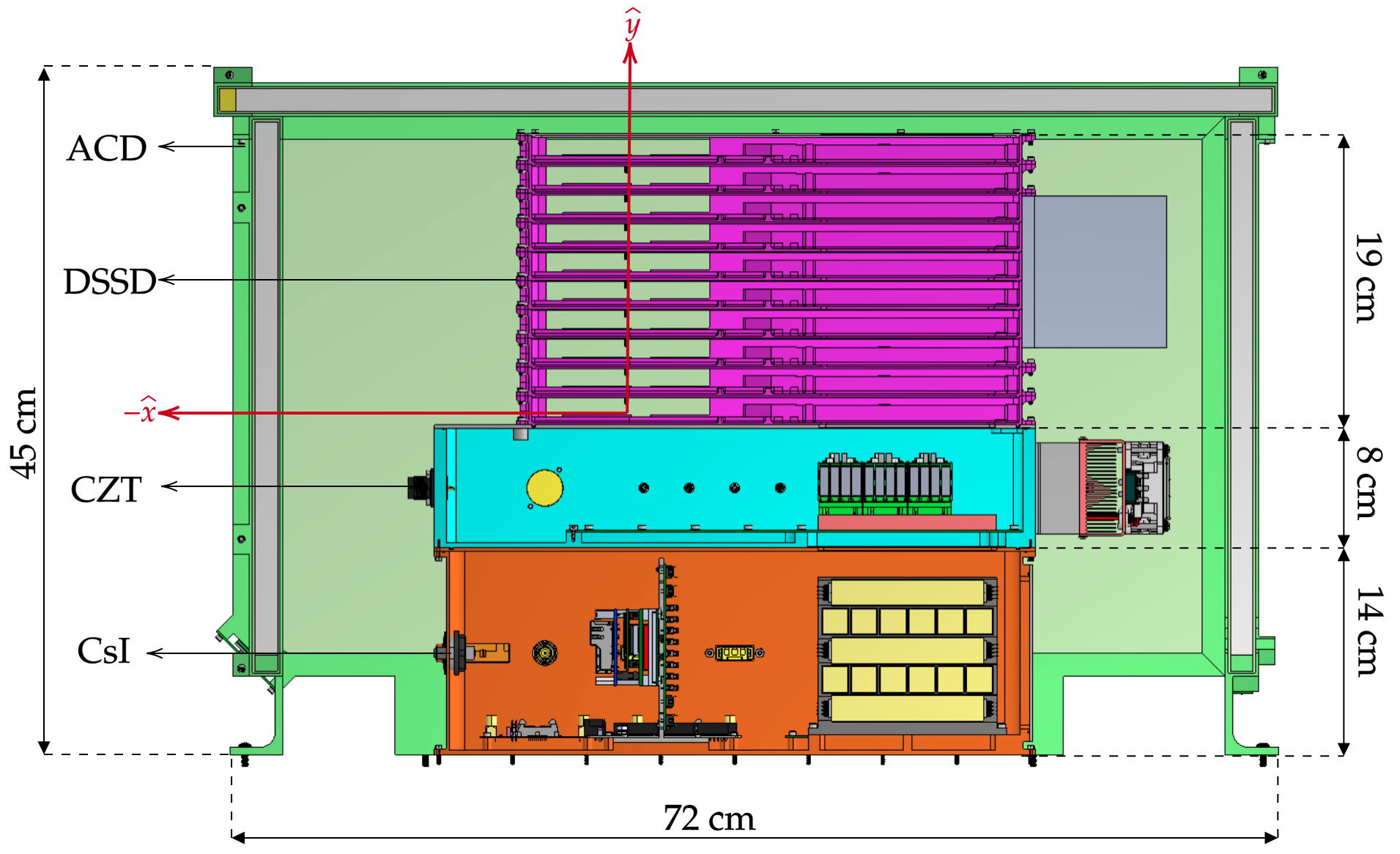}\hfill
\includegraphics[width=0.43\linewidth,angle=0]{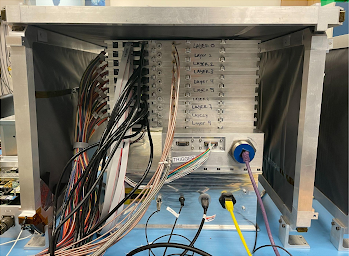}
\caption{\textit{Left:} CAD model of the ComPair balloon prototype. \textit{Right:} Picture of the ComPair instrument stack with one ACD panel removed.
\label{fig:scheme}}
\end{center}
\end{figure}

The ComPair instrument is a prototype next-generation gamma-ray observatory \cite{kierans20}. To demonstrate the instrument’s potential to measure photons that undergo Compton scattering or pair production processes, the ComPair team has developed a prototype that consists of all four AMEGO subsystems on a small scale and focused over a smaller energy range, 0.2 -- 25 MeV \citep{shy22}. Fig. \ref{fig:scheme} shows a computer aided design (CAD) and a photo of the ComPair instrument, where each subsystem is contained in an aluminum enclosure and can operate as a stand-alone detector. The DSSD Tracker \citep{griffin20} is at the top of the instrument, followed by the novel CZT calorimeter \citep{moiseev19}, and the CsI calorimeter \citep{woolf18} at the bottom; all of which are surrounded by an ACD that consists of five plastic scintillator panels. 

The DSSD tracker consists of 10 layers of 10 $\times$ 10 $\times$ 0.05 cm$^3$ Micron TTT13 detectors with 0.5 mm strip pitch. The DSSDs are AC-coupled and read out with IDEAS VATA460.3 application-specific integrated circuits (ASICs). 
The high-resolution virtual Frisch-grid CZT calorimeter \cite{moiseev19} provides an accurate measurement of the position and energy for Compton-scattered photons with an array of 6 $\times$ 6 $\times$ 20 mm$^3$ Redlen crystals that use custom AVG2 ASIC readout \cite{vernon19}. The CZT crystals fill an active area of 7.5 $\times$ 7.5 cm$^2$. 
The five layer hodoscopic CsI Calorimeter is lead by NRL and uses 30 CsI bars that are  1.7 $\times$ 1.7 $\times$ 10 cm$^3$ each and have a 0.25 -- 30 MeV energy range. The bars are read out with ONSemi Silicon photo-multipliers (SiPMs), and the IDEAS 64-channel ROSSPAD \cite{shy23}. 
The ACD panels are 1.5 cm thick EJ 208 plastic scintillator, with wavelength shifting bars along two edges that channel the light to ONSemi C-Series SiPMs in one corner.  
The ACD front-end is a duplication of the CsI readout.

ComPair's technology was inspired by the design of the Medium Energy Gamma-ray Astronomy \citep[MEGA;][]{2006ChJAS...6a.388B} instrument and the Tracking and Imaging Gamma-Ray Experiment \citep[TIGRE;][]{2006SPIE.6319E..19Z}, which flew on a balloon in 2007 and 2010.
For a more detailed description of the ComPair subsystems, we invite the reader to review \cite{shy22}. 

\subsection{Full balloon payload and peripherals}

\begin{figure}[h]
\begin{center}
\includegraphics[width=0.85\linewidth,angle=0]{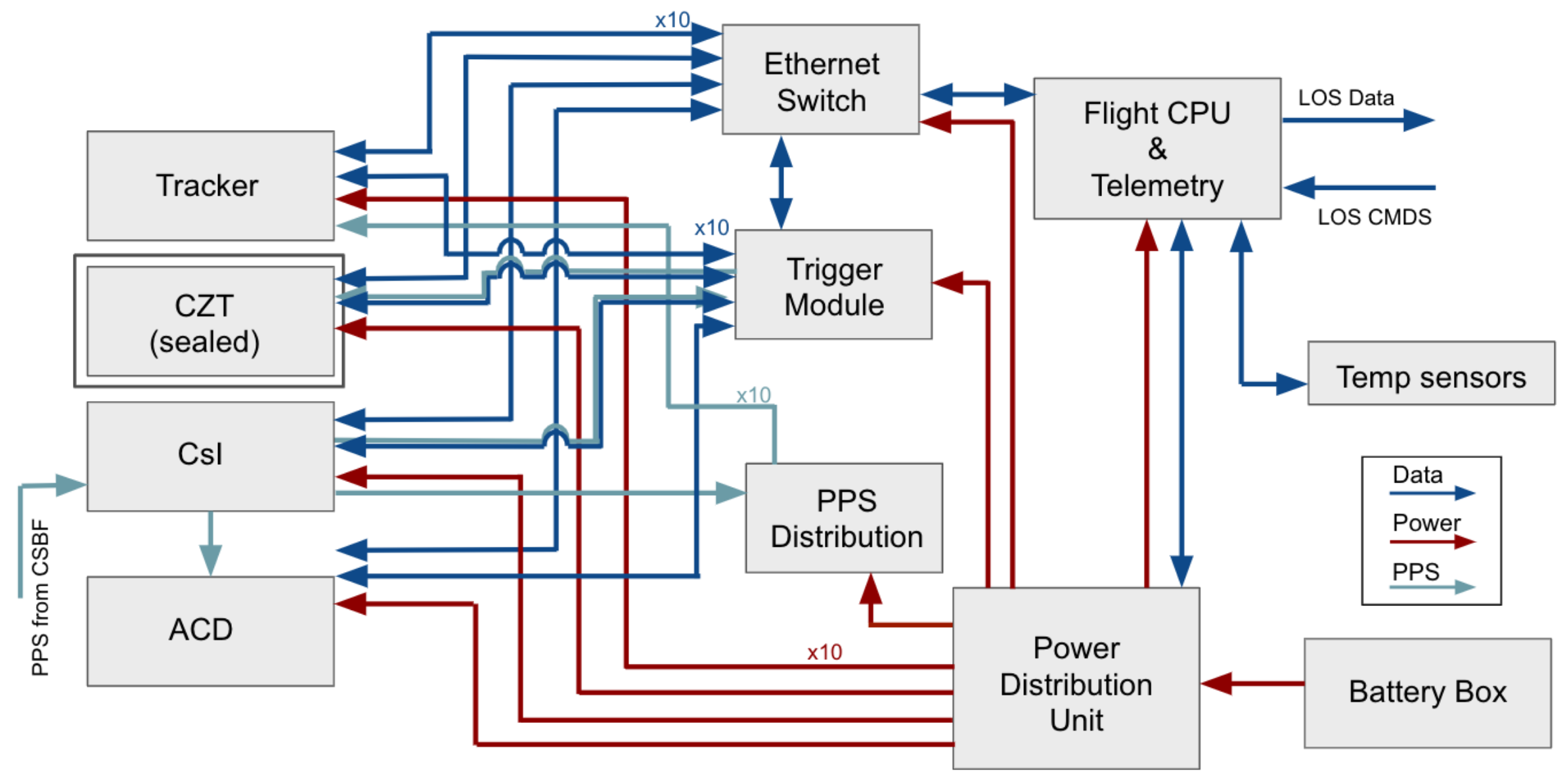}
\caption{ComPair instrument block diagram. 
The instrument consumes a total of 290 W and has a mass of 103 kg without batteries. 
\label{fig:payload}}
\end{center}
\end{figure}

\begin{figure}[h]
\begin{center}
\includegraphics[width=0.44\linewidth,angle=0]{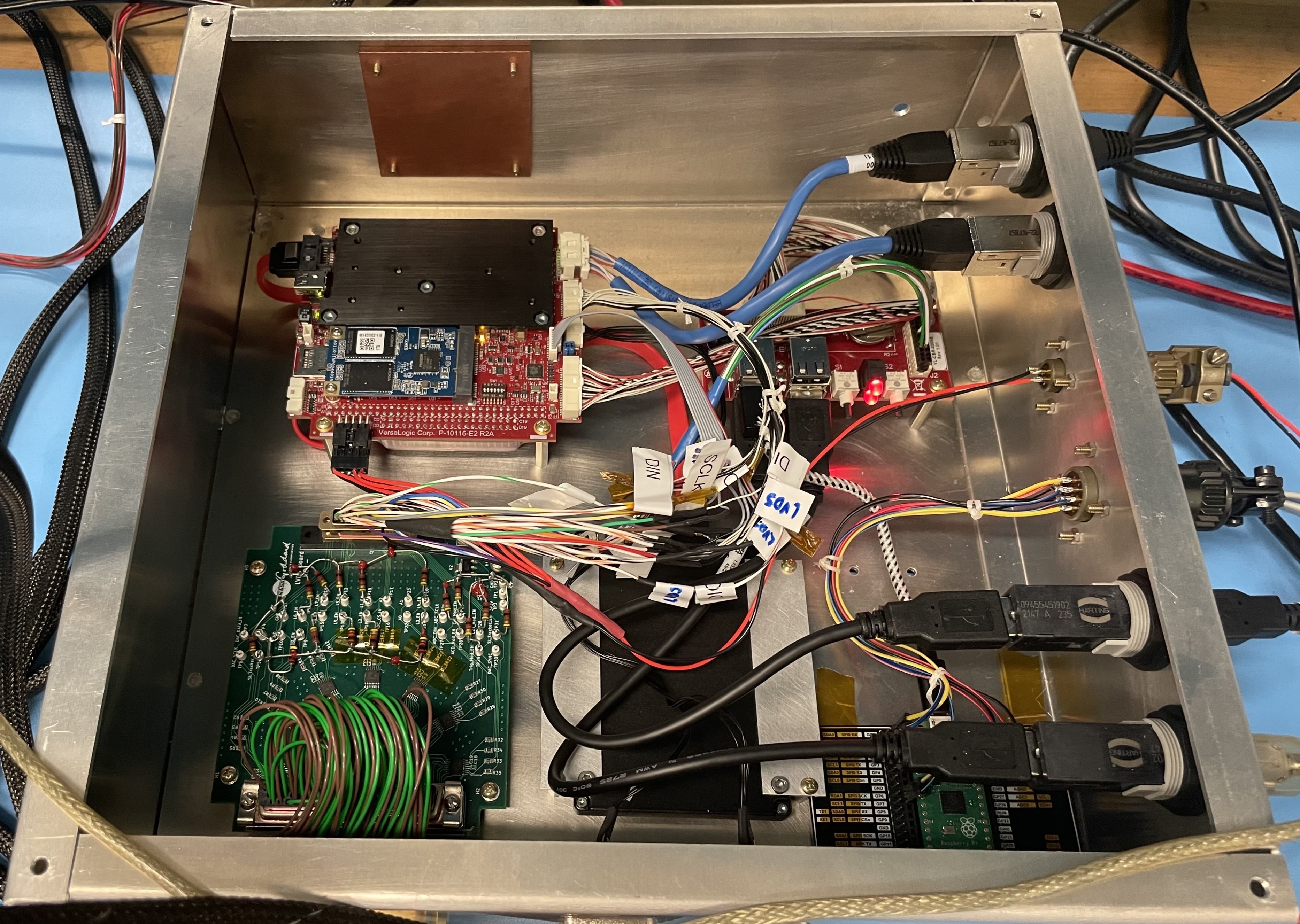}\hspace{6mm}
\includegraphics[width=0.44\linewidth,angle=0]{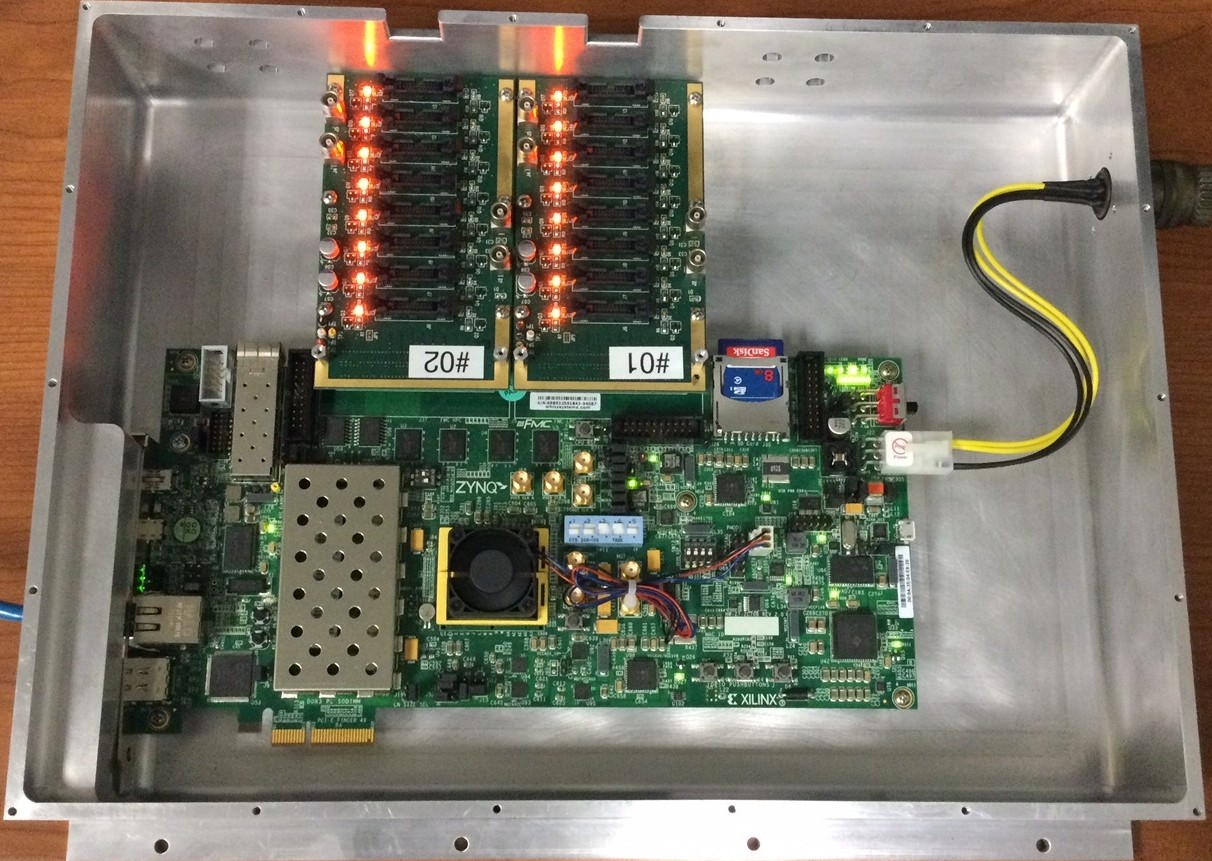}
\caption{\textit{Left:} The ComPair flight computer, or CPU. On the top left sits the Versalogic Baycat single board computer 
that provides digital input/output (I/O) connections through which the different subsystems are powered on. A paddle board is shown on its right which provide additional I/O connection. The BayCat's digital I/O pins toggle the low voltage differential signaling (LVDS) board shown on the bottom left part of the enclosure to enable and disable lines in the PDU. On the right of the LVDS board, sits the external backup hard drive. A Raspberry Pi Pico microcontroller (the bottom right) is used to monitor the temperature of the ComPair instrument subsystems and peripherals. \textit{Right:} The ComPair Trigger Module enclosure shows the ZC706 evaluation board with two trigger connector boards. 
\label{fig:peripherals}}
\end{center}
\end{figure}

A block diagram of the full ComPair balloon payload is shown in Fig. \ref{fig:payload}. The four ComPair subsystems communicate with a Trigger Module \cite{sasaki20} designed to check for coincidences between the subsystems within a programmable time window and tag events with an identification for off-line reconstruction (see Fig.~\ref{fig:peripherals} right). 
The Trigger Module uses a Xilinx ZC706 FPGA evaluation board, and provides 16 programmable coincidence modes and pre-scaling. 
The subsystems also receive a signal from a pulse-per-second (PPS) distribution unit to maintain synchronization. The ComPair power distribution unit (PDU) distributes power to all of the subsystems and peripherals, and will be powered by 28~V batteries during the balloon flight. 

The flight computer, or Central Processing Unit (CPU), was developed by LANL and NASA GSFC and drives and manages the data acquisition for the different subsystems (see Fig. \ref{fig:peripherals} left). Its architecture is based on a Versalogic BayCat single board computer and provides in-flight commanding and data downlink. The computer uses an Elixir Open Telecom Platform to supervise subsystem-specific long-running processes, manages data collection for each subsystem, controls the PDU and readout thermometry. The CPU connects to the Columbia Scientific Balloon Facility (CSBF) miniature Science Instrument Package (SIP) to downlink housekeeping and data during the flight.  

\begin{figure}[h]
\begin{center}
\includegraphics[width=0.54\linewidth,angle=0]{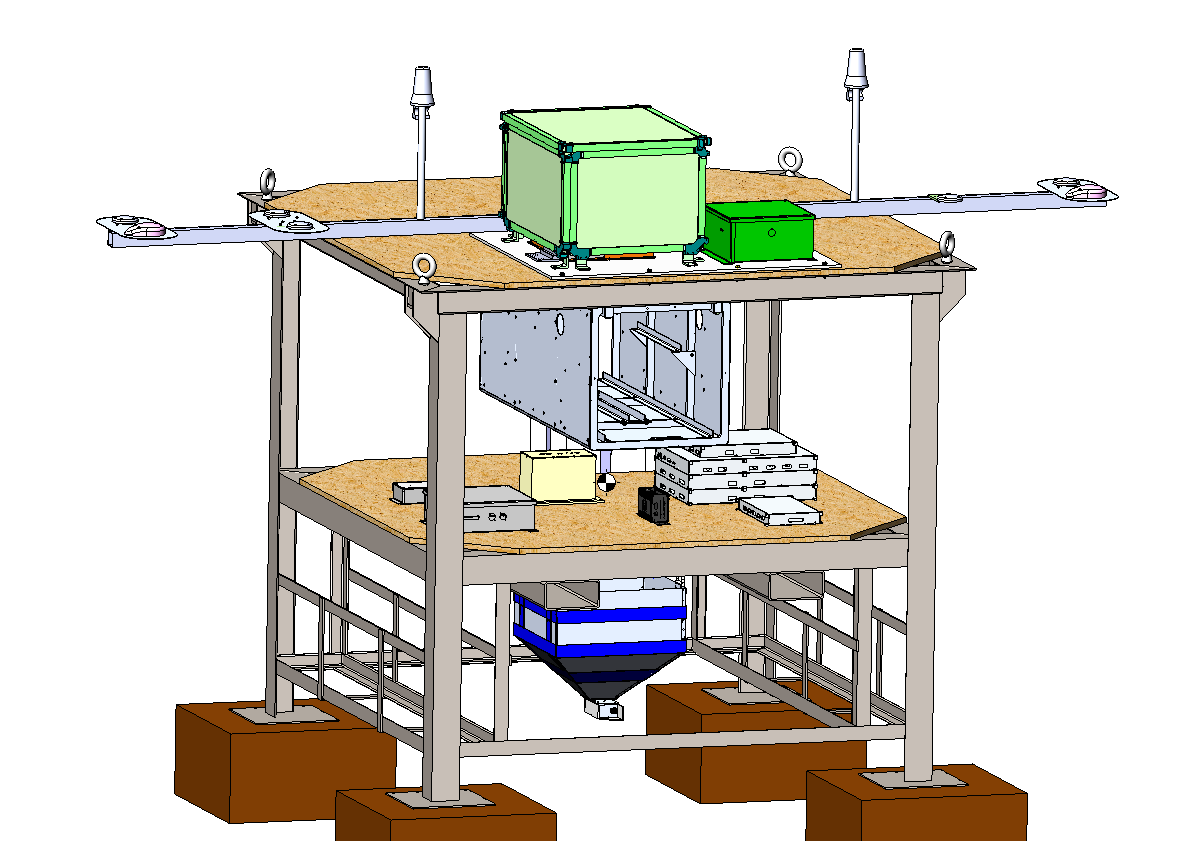}
\includegraphics[width=0.43\linewidth,angle=0]{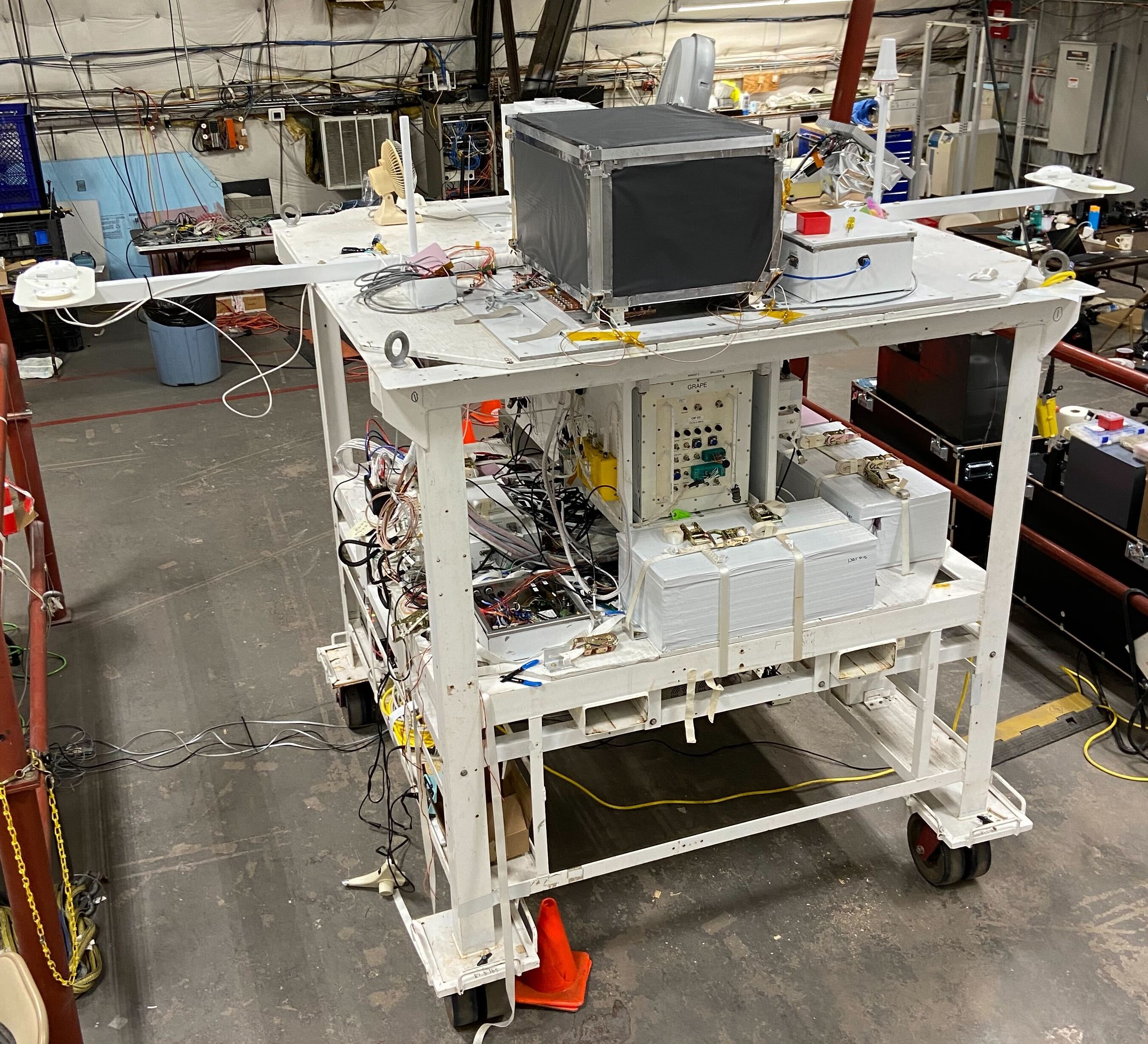}
\caption{CAD (\textit{left}) and photo (\textit{right}) of the actual ComPair payload integrated on the CSBF gondola.
The GRAPE and IRCSP instruments sit next to the ComPair base plate on the top level of the gondola.
\label{fig:gondola}}
\end{center}
\end{figure}

The ComPair instrument is positioned on the top deck of the CSBF Iron Maiden gondola for flight, as shown in Figure \ref{fig:gondola}. The ACD data acquisition enclosure and Trigger Module will sit next to the instrument stack on top of an aluminum baseplate. The PDU, CPU and Ethernet switch are positioned in the middle deck beneath the suspended miniature SIP.

\section{ComPair prototype tests}

The ultimate functionality test of the ComPair prototype will be the balloon flight scheduled to take place in August 2023 from Fort Sumner, New Mexico. 
The goal is to measure the gamma-ray albedo background at float altitudes 
$\sim40$ km above Earth's surface, and prove the functionality of the prototype telescope in a space-like environment. 
We aim for a duration of six hours at float altitude and will collect real-time telemetry of basic housekeeping and event rates. The full instrument data will be recovered from the hard drives after the flight.
ComPair will fly on CSBF’s gondola with another MeV instrument, the Gamma-RAy Polarimeter Experiment \citep[GRAPE;][]{2022SPIE12241E..0GO}, and with the InfraRed Channeled Spectro-Polarimeter \citep[IRCSP;][]{2022SPIE12112E..0LH}. 

\subsection{Thermal vacuum test}

\begin{figure}[h]
\begin{center}
\includegraphics[width=\linewidth,angle=0]{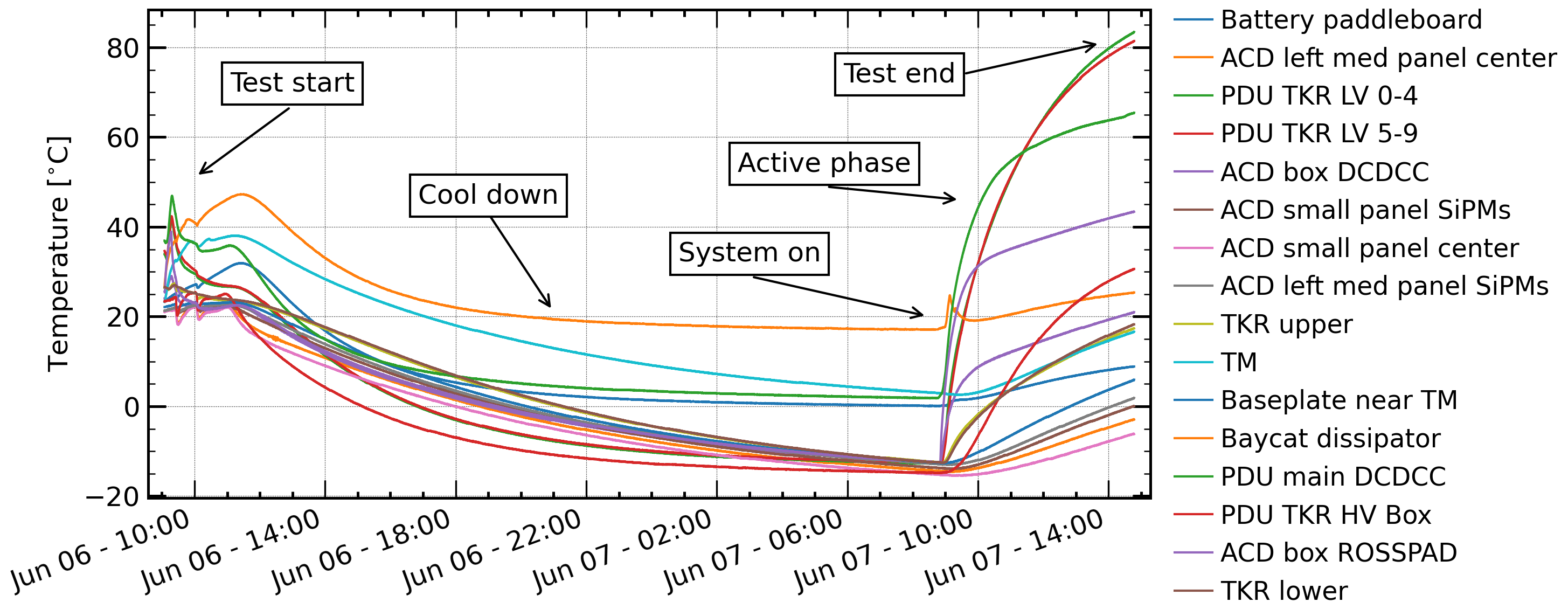}
\caption{Temperatures measured during TVAC. The legend on the right indicates where the thermometers were located on the ComPair instrument subsystems and peripherals. The PDU temperatures exhibited the most dramatic changes, which was later addressed by adding copper heat straps within the PDU enclosure. 
\label{fig:tvac}}
\end{center}
\end{figure}

In preparation for the balloon flight, the full ComPair payload underwent thermal vacuum (TVAC) testing at NASA GSFC's Integration and Testing facility in Chamber 239. The goal of this test was to benchmark thermal models for the balloon flight and test the functionality of the PDU and CPU in flight conditions. 
The chamber operated at 5 Torr and the walls were maintained at -20$^{\circ}$C for the duration of the test. As shown in Fig.~\ref{fig:tvac}, the chamber was initially cooled with the instrument powered off for the first 24 hours to equilibrate. 
The instrument was then fully powered on for a 7 hour test. The measured temperatures of each component were compared to a Thermal Desktop model with the same configuration, and small modifications were made to the thermal model to better match the TVAC data. This allowed for a higher fidelity prediction of the flight temperatures. 

\subsection{Calibrations and performance test}

ComPair was tested at the Triangle Universities Nuclear Laboratory’s High Intensity Gamma-ray Source (HIGS) in April 2022. It provided a mono-energetic gamma-ray beam of 2 to 25 MeV; and a flux of $\sim 1000$ ph/s, allowing for tests of the instrument in the pair regime not accessible through laboratory radioactive sources. The Tracker, CZT and CsI successfully operated with the Trigger Module; however, the ACD was not integrated. Preliminary analysis results from the beam test are found in \cite{shy22}.  

\begin{figure}[h]
\begin{center}
\includegraphics[width=0.32\linewidth,angle=0]{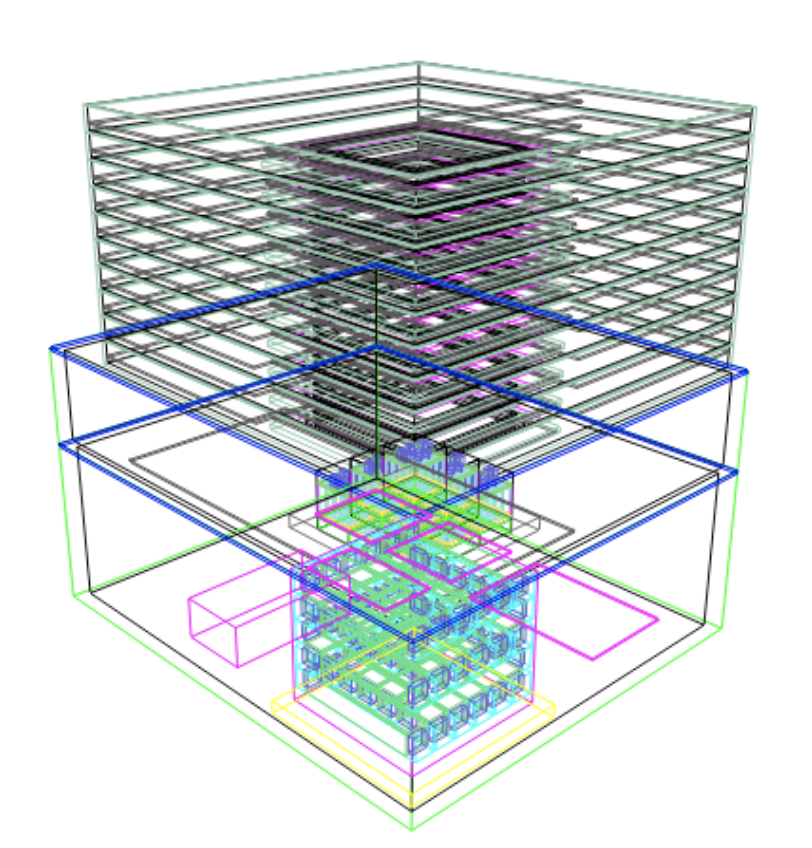}\hspace{20mm}
\includegraphics[width=0.4\linewidth,angle=0]{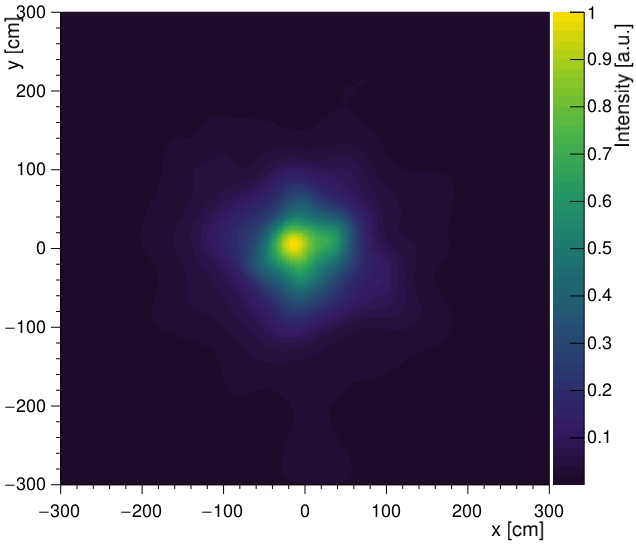}\vspace{1mm}
\includegraphics[width=0.49\linewidth,angle=0]{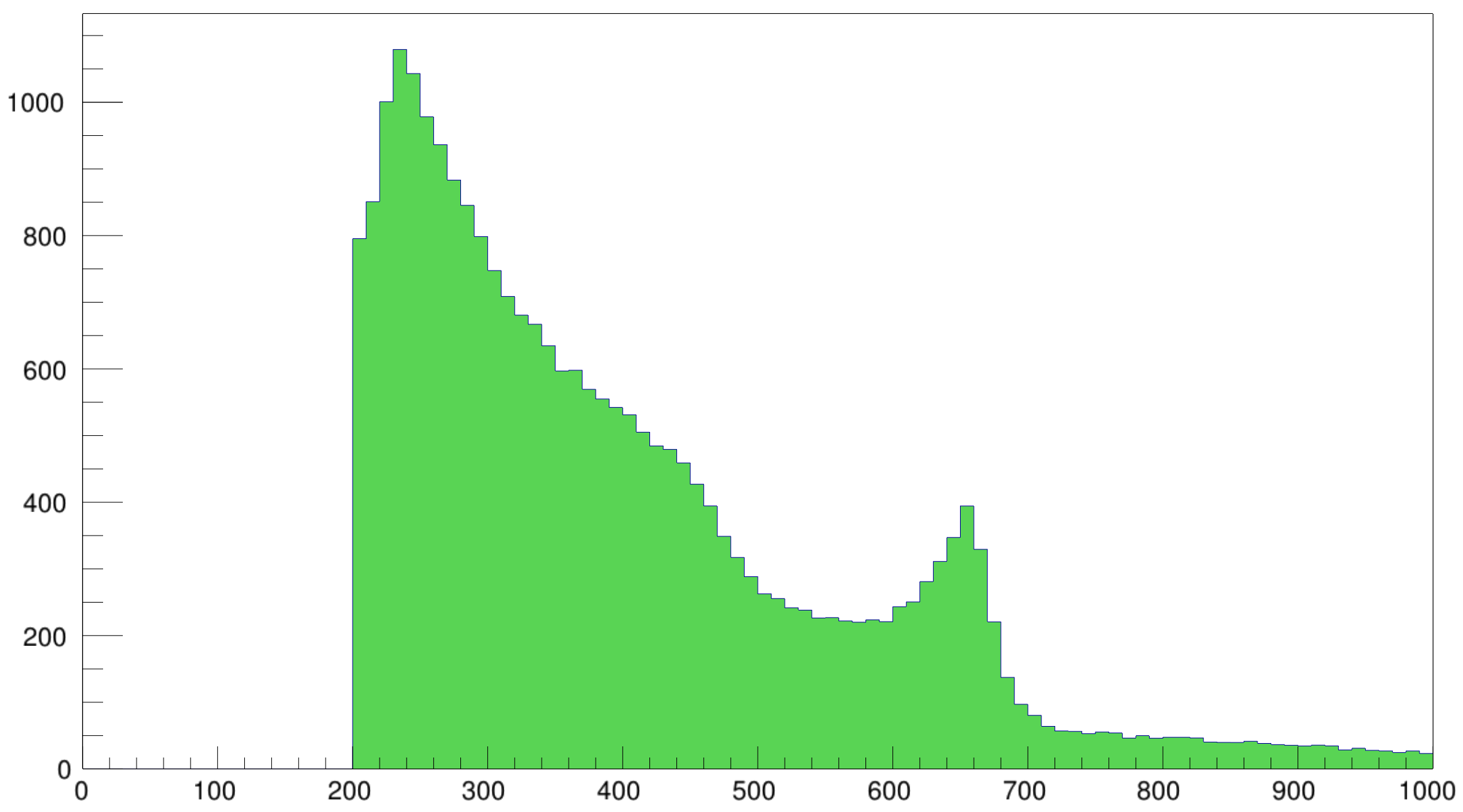}\hspace{4mm}
\includegraphics[width=0.44\linewidth,angle=0]{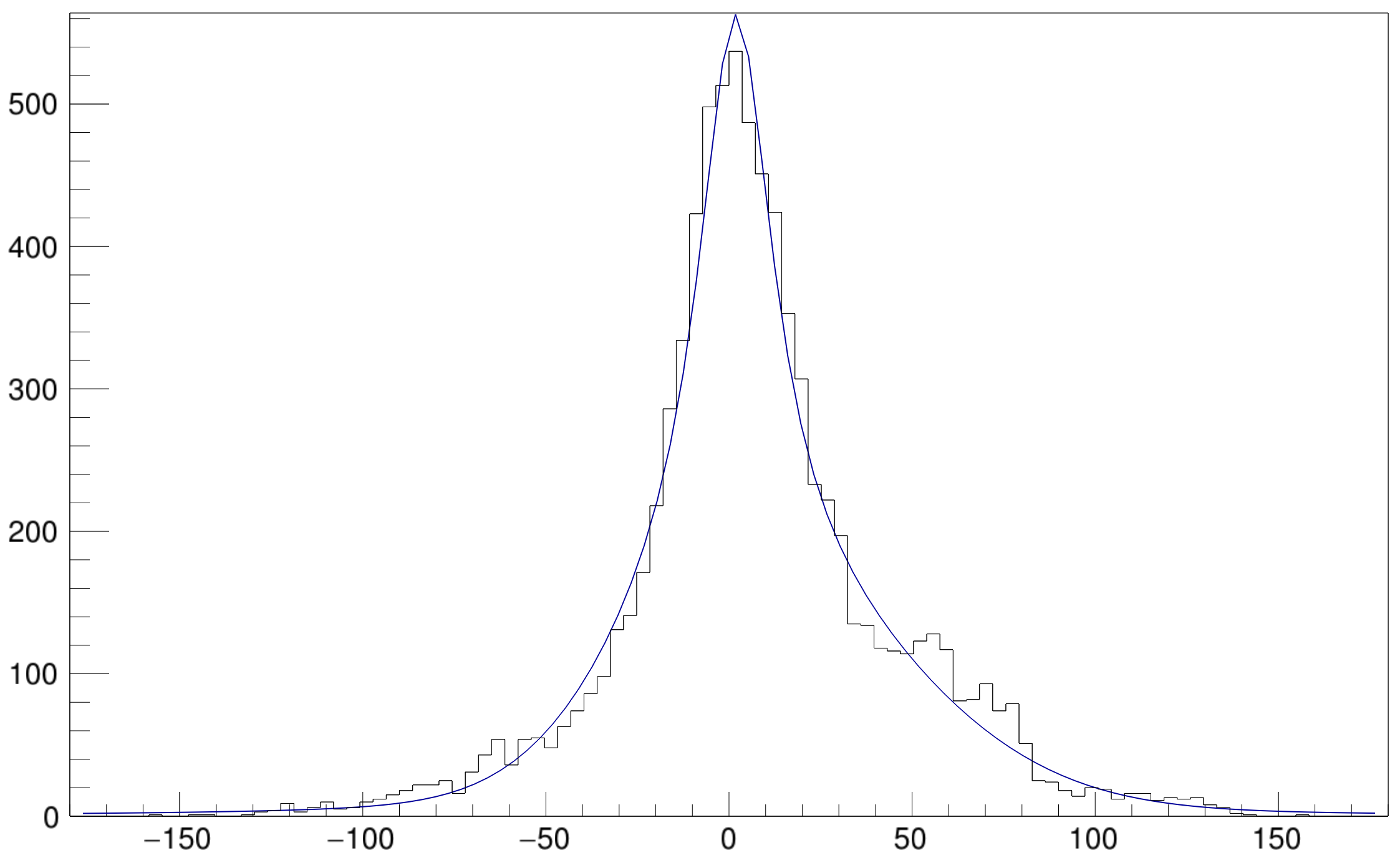}
\put(-290,100){Cs-137 (662 keV)} \put(-290,85){PRELIMINARY} 
\put(-427,57){\rotatebox{90}{Counts / keV}} \put(-265,-10){Energy [keV]}
\put(-200,80){\rotatebox{90}{Counts}} \put(-115,-10){ARM - Compton cone [$^{\circ}$]}
\caption{\textit{Top left:} MEGAlib mass model of ComPair. \textit{Top right:} Event reconstructed image of calibration measurements with a Cs-137 (662 keV) radioactive source, for the integrated ComPair instrument. \textit{Bottom left:} Corresponding event spectrum after reconstruction. \textit{Bottom right:} ARM from the Compton cone. 
\label{fig:calib}}
\end{center}
\end{figure}

Full system calibrations and performance tests in the Compton regime have been performed with laboratory radioactive sources both at NASA GSFC and in Fort Sumner, NM.  
Fig. \ref{fig:calib} shows a preliminary spectrum, image, and angular resolution measurement (ARM) after event reconstruction. A Cesium 137 radioactive source was located $\sim3$~m above ComPair for a 3~hour integration time. The high energy continuum in the spectrum is due to muons and chance coincidence events. A preliminary energy resolution of $\Delta E/E = 12\%$ full width at half maximum (FWHM) was found; however, it is expected that this value will improve after optimized subsystem calibrations. The FWHM of the ARM distribution, i.e. the angular resolution, was found to be 33$^{\circ}$, where only events with a total energy deposit between 600 and 680 keV were selected. 
The ComPair pipeline performs simulations, event reconstruction, and analysis with MEGAlib \cite{zoglauer06} (see mass model on the top left-hand side of Fig. \ref{fig:calib}). The full system measurements are still underway and performance numbers are preliminary and expected to improve with better calibrations and event selections.

\section{Summary}

The ComPair telescope has undergone multiple functionality and performance tests, including a beam test, thermal vacuum tests, thermal modeling, and calibration and performance tests with radioactive sources. The results from the balloon flight and calibration campaign will appear in a forthcoming publication. 
At the time of writing, the team is at Fort Sumner, New Mexico, taking care of the last details before the ComPair balloon flight, scheduled to take place in August 2023. 

\section{Acknowledgments}
This material is based upon work supported by NASA under award number 80GSFC21M0002.
This work was sponsored by NASA-APRA (NNH21ZDA001N-APRA, NNH14ZDA001N-APRA and NNH15ZDA001N-APRA). A. W. Crosier and T. Caligure were sponsored by the Office of Naval Research NREIP Program.

%
%
%


\begin{thebibliography}{99}\vspace{-1mm}
\setlength{\itemsep}{-.7mm} 
\bibitem{1988Natur.331..416M} S. Matz et al. 
1988, Nature, \href{https://ui.adsabs.harvard.edu/abs/1988Natur.331..416M/abstract}{331, 416}.
\bibitem{2018ApJ...861...85A} M. Ajello et al. 
2018, ApJ, \href{https://ui.adsabs.harvard.edu/abs/2018ApJ...861...85A/abstract}{861, 85}. 
\bibitem{IceCube:2018cha} The IceCube Collaboration 
2018, Science, \href{https://ui.adsabs.harvard.edu/abs/2018Sci...361..147I/abstract}{361, 6398, 147--151}.
\bibitem{1993ApJS...86..657S} V. Schoenfelder et al. 1993, ApJS, \href{https://ui.adsabs.harvard.edu/abs/1993ApJS...86..657S/abstract}{86, 657}.
\bibitem{kierans20} C. Kierans et al. 
2020, Proceedings of the SPIE, \href{https://ui.adsabs.harvard.edu/abs/2020SPIE11444E..31K/abstract}{11444, 1144431}.
\bibitem{mcenery19} J. McEnery, et al. 
2019, \href{https://asd.gsfc.nasa.gov/amego/files/AMEGO_Decadal_RFI.pdf}{PDF}.
\bibitem{2009ApJ...697.1071A} W. B. Atwood et al. 
2009, ApJ, \href{https://ui.adsabs.harvard.edu/abs/2009ApJ...697.1071A/abstract}{697, 1071}. 
\bibitem{2005SSRv..120..165B} D. Burrows et al. 
2005, SSR, \href{https://ui.adsabs.harvard.edu/abs/2005SSRv..120..165B/abstract}{120, 165}.
\bibitem{2019BAAS...51g..98T} J. Tomsick et al. 
2019, \href{https://ui.adsabs.harvard.edu/abs/2019BAAS...51g..98T/abstract}{BAAS}.
\bibitem{shy22} D. Shy et al. 
2022, Proceedings of the SPIE \href{https://ui.adsabs.harvard.edu/abs/2022SPIE12181E..2GS/abstract}{12181, 121812G}.
\bibitem{griffin20} S. Griffin et al. 
2020, Proceedings of the SPIE \href{https://ui.adsabs.harvard.edu/abs/2020SPIE11444E..34G/abstract}{11444, 1144434}.
\bibitem{moiseev19} A. Moiseev et al. 
2019, \href{https://pos.sissa.it/358/584/pdf}{ICRC Proceedings}.
\bibitem{woolf18} R. Woolf 
2018, IEEE \href{https://ui.adsabs.harvard.edu/abs/2019LPICo2135.5016W/abstract}{NSS MIC}.
\bibitem{vernon19} E. Vernon et al. 
2019, NIM A \href{https://ui.adsabs.harvard.edu/abs/2019NIMPA.940....1V/abstract}{940}.
\bibitem{shy23} D. Shy, et al. 
2023, \href{https://ui.adsabs.harvard.edu/abs/2023arXiv230711177S/abstract}{arXiv:2307.11177}. 
\bibitem{2006ChJAS...6a.388B} P. Bloser et al. 
2006, Chinese Journal of Astronomy and Astrophysics Supplement, \href{https://ui.adsabs.harvard.edu/abs/2006ChJAS...6a.388B/abstract}{6, 388}.
\bibitem{2006SPIE.6319E..19Z} A. Zych et al. 
2006, Proceedings of the SPIE, \href{https://ui.adsabs.harvard.edu/abs/2006SPIE.6319E..19Z/abstract}{6319, 631919}.
\bibitem{sasaki20} M. Sasaki 
2020,  Proceedings of the SPIE \href{https://ui.adsabs.harvard.edu/abs/2020SPIE11444E..6AS/abstract}{11444, 114446A}.
\bibitem{2022SPIE12241E..0GO} M. O{\~n}ate Melecio et al. 2022, 
Proceedings of the SPIE, \href{https://ui.adsabs.harvard.edu/abs/2022SPIE12241E..0GO/abstract}{12241, 122410G}. 
\bibitem{2022SPIE12112E..0LH} K. Hart Shanks et al. 
2022, Proceedings of the SPIE, \href{https://ui.adsabs.harvard.edu/abs/2022SPIE12112E..0LH/abstract}{12112, 121120L}. 
\bibitem{zoglauer06} A. Zoglauer et al. 
2006, New Astro. Rev., \href{https://ui.adsabs.harvard.edu/abs/2006NewAR..50..629Z/abstract}{50}.

\end{thebibliography}
\end{document}